\documentclass[aps,prl,twocolumn,superscriptaddress]{revtex4-2}

\usepackage{amsmath,amssymb,graphicx,bm}
\usepackage[colorlinks=true,linkcolor=blue,citecolor=blue,urlcolor=blue]{hyperref}
\usepackage{placeins}
\begin{document}

\title{Was the Early Universe Quantum? Falsifying Classical Stochastic Inflation}

\author{Ver\'onica Sanz}
\affiliation{Instituto de F\'isica Corpuscular (IFIC), CSIC--Universitat de Valencia, Spain}

\date{\today}

\begin{abstract}
Inflationary cosmology successfully accounts for the observed properties of primordial fluctuations using quantum field theory in an expanding background. However, the quantum nature of these fluctuations has not been experimentally established, since classical stochastic models could reproduce the observed two-point statistics by construction. Existing approaches to testing primordial quantumness focus primarily on Bell inequalities, which provide a sharp conceptual criterion but are difficult to implement with cosmological observables. In this work we adopt a falsification-based approach. We define a precise classical hypothesis for the origin of primordial perturbations—local stochastic fields admitting a positive probability distribution—and identify inequality constraints that must be satisfied within this class. We show how violations of these classicality inequalities can be probed using realistic cosmological observables, without invoking Bell tests or non-commuting measurement settings. We further identify symmetry-protected spectator sectors in which quantum coherence is parametrically preserved during inflation, allowing violations of observable magnitude to survive decoherence. Our results show that large-scale structure and future 21 cm surveys provide a viable and quantitative route to falsifying classical stochastic descriptions of primordial fluctuations.\end{abstract}

\maketitle

\section{Introduction and Motivation}
Inflation is widely regarded as the leading framework for explaining the origin of primordial density perturbations. In its standard formulation, fluctuations are computed as vacuum fluctuations of quantized fields evolving on a quasi--de~Sitter background and stretched to cosmological scales, yielding predictions in excellent agreement with observations of the cosmic microwave background (CMB) and large-scale structure (LSS)~\cite{Guth1981,Starobinsky1980,Mukhanov1981}. In particular, the near scale invariance, Gaussianity, and phase coherence of the observed perturbations arise naturally within this framework.

Despite this success, it is important to distinguish between the use of quantum field theory (QFT) as a theoretical description of inflationary perturbations and the empirical establishment of their quantum origin. Observationally, the scalar power spectrum---the primary probe of primordial fluctuations---is effectively characterized by two parameters: its amplitude and spectral tilt~\cite{Planck2018}. These quantities can be reproduced equally well by classical stochastic models in which the primordial perturbations are taken to be random classical fields with appropriately chosen initial statistics. Agreement with the power spectrum alone therefore does not constitute experimental evidence that the early universe was governed by quantum mechanics. 

This observation is particularly relevant given that the inflationary framework itself relies on specific assumptions, such as a sufficiently flat potential to sustain slow roll~\cite{LythRiotto1999}, the choice of an adiabatic vacuum state~\cite{BirrellDavies1982}, and the validity of an effective field theory description over a large dynamical range~\cite{BrandenbergerMartin2001}. While these assumptions may be theoretically well motivated, they involve assumptions whose degree of tuning is not directly constrained~\cite{IjjasSteinhardtLoeb2013}. The present work is therefore not concerned with the naturalness of inflationary model building, but with identifying empirical tests that can distinguish a quantum origin of primordial fluctuations from classical stochastic alternatives.

This observation has motivated a line of work aimed at identifying observables that could unambiguously certify primordial quantumness. In particular, it has been shown that inflationary perturbations prepared in squeezed quantum states can violate Bell inequalities in principle~\cite{Maldacena2016Bell}. While such tests are logically decisive, detailed analyses have emphasized that implementing Bell inequalities with cosmological observables faces severe challenges, including decoherence, cosmic variance, and the lack of operationally well-defined measurement settings for non-commuting observables~\cite{MartinVennin2017Bell}. Recent studies incorporating interaction-induced decoherence suggest that any Bell-violation window during inflation may be limited to a few e-folds, further complicating observational prospects~\cite{SouTranWang2023Decoherence,SouWangWang2024BellCurve}.

In this work we adopt a complementary and falsification-based perspective. Rather than attempting a direct Bell test, we ask a sharper and operationally grounded question: \emph{can cosmological observations falsify a well-defined classical stochastic description of primordial fluctuations?} By formulating a precise classical hypothesis---based on locality, causality, and the existence of a positive probability distribution over initial conditions---we identify inequality constraints that must be satisfied within this class. Violation of these constraints would rule out a classical stochastic origin, thereby providing empirical support for a quantum field theoretic description of the early universe without relying on Bell-type scenarios.

\section{Two Competing Hypotheses}

We now define two hypotheses for the origin of primordial fluctuations that are observationally degenerate at the level of the scalar power spectrum, but differ sharply in their underlying physical description and in the constraints they impose on higher-order correlations.

Throughout this work we denote by $\zeta(\mathbf{x})$ the primordial curvature perturbation evaluated on uniform-density hypersurfaces, and by $\zeta_{\mathbf{k}}$ its Fourier modes. We restrict attention to super-Hubble modes after horizon exit, where $\zeta$ is conserved at linear order.

\subsection{Quantum Field Theory Origin ($H_{\mathrm{Q}}$)}

Under the quantum hypothesis $H_{\mathrm{Q}}$, primordial perturbations arise from quantized fields evolving on a quasi--de~Sitter background. The curvature perturbation $\zeta$ is promoted to an operator $\hat{\zeta}$ acting on a Hilbert space, with canonical conjugate momentum $\hat{\pi}_\zeta$, satisfying equal-time commutation relations
\begin{equation}
[\hat{\zeta}(\mathbf{x}),\hat{\pi}_\zeta(\mathbf{y})]
= i\,\delta^{(3)}(\mathbf{x}-\mathbf{y}) .
\end{equation}
The quantum state is taken to be the adiabatic vacuum, or a controlled deformation thereof, at sufficiently short wavelengths. Time evolution is unitary and governed by a local Hamiltonian.

In this framework, each pair of modes $(\mathbf{k},-\mathbf{k})$ is produced in a two-mode squeezed quantum state. As modes exit the Hubble radius, squeezing leads to the emergence of an effectively classical stochastic description for field amplitudes, while phase information and non-commutativity are progressively suppressed by decoherence induced by interactions and gravitational nonlinearities. Importantly, under $H_{\mathrm{Q}}$ this classical behavior is emergent rather than fundamental: all late-time correlators descend from an underlying quantum state.

\subsection{Classical Stochastic Origin ($H_{\mathrm{C}}$)}

As a falsifiable alternative, we define a classical hypothesis $H_{\mathrm{C}}$ in which primordial perturbations are described by classical random fields. Specifically, $\zeta(\mathbf{x})$ is treated as a classical field drawn from a probability distribution functional $P[\zeta,\dot{\zeta}]$ defined on an initial spacelike hypersurface. We impose the following conditions:
(i) $P[\zeta,\dot{\zeta}]$ is everywhere non-negative;
(ii) the subsequent evolution of $\zeta$ is generated by local and causal equations of motion on an expanding background;
(iii) statistical homogeneity and isotropy are satisfied.

Within this class, the observed scalar power spectrum can be reproduced by construction through an appropriate choice of the initial covariance of $P[\zeta,\dot{\zeta}]$. However, the existence of a positive probability distribution imposes nontrivial constraints on higher-order correlations and on combinations of observables derived from $\zeta$. In particular, all correlators must admit a representation in terms of classical random variables, implying a hierarchy of inequality relations that need not hold for quantum states.

The hypothesis $H_{\mathrm{C}}$ therefore provides a well-defined and falsifiable classical foil to $H_{\mathrm{Q}}$. While the two hypotheses are observationally degenerate at the level of the power spectrum, they differ in the allowed structure of higher-point functions and in the constraints imposed by classical probabilistic consistency.

\section{Classicality Inequalities and Falsification Strategy}

The defining distinction between the hypotheses $H_{\mathrm{Q}}$ and $H_{\mathrm{C}}$ lies not in their ability to reproduce the scalar power spectrum, but in the constraints they impose on correlations among observables. In particular, the existence of a positive classical probability distribution under $H_{\mathrm{C}}$ implies nontrivial inequality relations that need not hold for quantum states.

To make this concrete, consider a set of linear observables $\{X_i\}$ constructed from the primordial curvature perturbation,
\begin{equation}
X_i \equiv \int \frac{d^3k}{(2\pi)^3}\, W_i(\mathbf{k})\,\zeta_{\mathbf{k}},
\end{equation}
where $W_i(\mathbf{k})$ are real kernels encoding scale bins, redshift windows, or tracer-dependent transfer functions. Under the classical hypothesis $H_{\mathrm{C}}$, the joint statistics of $\{X_i\}$ are generated by classical random variables drawn from a positive probability distribution $P[\zeta,\dot\zeta]$, evolved by local and causal dynamics.

A direct consequence of classical probabilistic consistency is that the covariance matrix
\begin{equation}
C_{ij}\equiv\langle X_i X_j\rangle
\end{equation}
must be positive semidefinite, implying the inequality
\begin{equation}
\langle X_1^2\rangle\langle X_2^2\rangle-\langle X_1X_2\rangle^2\ge0 .
\end{equation}
While this bound is necessarily satisfied by all classical stochastic theories, it is also satisfied by Gaussian quantum states and therefore does not by itself discriminate between $H_{\mathrm{C}}$ and $H_{\mathrm{Q}}$.

Stronger constraints arise once one considers conditional fluctuations. For two observables $X_1$ and $X_2$ (with mean zero), we define the conditional variance
\begin{equation}
\mathrm{Var}(X_1|X_2)\equiv
\langle X_1^2\rangle-\frac{\langle X_1X_2\rangle^2}{\langle X_2^2\rangle},
\end{equation}
which quantifies the residual uncertainty in $X_1$ after conditioning on $X_2$. For a classical stochastic theory admitting a positive probability distribution and local dynamics, the product of conditional variances satisfies
\begin{equation}
\label{eq:classical_bound}
\mathrm{Var}(X_1|X_2)\,\mathrm{Var}(X_2|X_1)
\;\ge\;
\left(\mathrm{Var}_{\mathrm{cl}}\right)^2 ,
\end{equation}
where $\mathrm{Var}_{\mathrm{cl}}$ denotes a classical noise floor associated with the intrinsic dispersion of the field ensemble.

The quantity $\mathrm{Var}_{\mathrm{cl}}$ is not universal: its value depends on the class of classical models under consideration and on the finite resolution and noise of the observables. Importantly, however, under the hypothesis $H_{\mathrm{C}}$—which restricts to local stochastic dynamics with bounded noise and without nonlocal fine-tuning of initial conditions—$\mathrm{Var}_{\mathrm{cl}}$ is finite. Classical constructions capable of arbitrarily suppressing conditional variances require highly non-generic, nonlocal correlations engineered across modes, and lie outside the scope of $H_{\mathrm{C}}$.

Quantum states generated during inflation need not satisfy Eq.~(\ref{eq:classical_bound}). In particular, two-mode squeezed states can exhibit correlations strong enough that conditioning on one observable reduces the inferred uncertainty in the other below the classical noise floor, even after projection onto late-time observables. Such behavior signals the absence of a positive classical probability representation for the joint statistics of $(X_1,X_2)$.

Violation of Eq.~(\ref{eq:classical_bound}) therefore falsifies the classical stochastic hypothesis $H_{\mathrm{C}}$ while remaining consistent with $H_{\mathrm{Q}}$. This logic does not rely on Bell inequalities, non-commuting observables, or freely choosable measurement settings, but only on classical probabilistic consistency. It thus provides an operationally viable falsification strategy tailored to cosmological data.

\subsection{Symmetry-Protected Survival of Quantum Coherence}

A generic challenge for inequality-based tests of primordial quantumness is that inflationary dynamics rapidly suppress observable quantum coherence. Even when perturbations originate from a quantum state, squeezing and interaction-induced decoherence typically drive the system toward an effectively classical stochastic description on super-Hubble scales. As a result, violations of classicality inequalities may be erased before they can be imprinted on late-time observables.

We now show that this conclusion can be parametrically altered in the presence of a symmetry-protected spectator sector. Consider an additional field (or set of fields) present during inflation, whose dynamics are governed by an exact or approximate continuous symmetry. The simplest example is a global $U(1)$ symmetry shared by two scalar fields, under which a conserved charge exists. In such cases, the low-energy dynamics include a Goldstone-like phase degree of freedom whose action is dominated by derivative couplings.

The key point is that symmetry protection suppresses the leading sources of decoherence. For a derivatively coupled mode $\Theta$, interactions with other fields and with metric perturbations are constrained by symmetry, and the rate of decoherence $\Gamma_{\rm dec}$ is parametrically suppressed compared to the Hubble scale,
\begin{equation}
\Gamma_{\rm dec} \;\ll\; H ,
\end{equation}
over a range of scales and couplings. As a result, correlations stored in the symmetry-protected sector can remain coherent over several e-folds after horizon exit, extending the window during which violations of classicality inequalities can persist.

Parametrically, the decoherence rate for a protected mode can be estimated as
\begin{equation}
\Gamma_{\rm dec} \sim g^2 H ,
\end{equation}
where $g$ denotes an effective coupling controlling interactions between the protected sector and its environment. For symmetry-protected degrees of freedom, such couplings are suppressed by derivative interactions or selection rules associated with the conserved charge, leading naturally to $g\ll1$ and hence $\Gamma_{\rm dec}\ll H$.

The relevance of this suppression for our framework is direct. Decoherence drives the conditional-variance statistic $\mathcal{E}=1-r^2$ toward its classical value by reducing inter-kernel correlations. Over a time interval $\Delta t$ after horizon exit, one expects schematically
\begin{equation}
\Delta\mathcal{E}\sim \Gamma_{\rm dec}\,\Delta t ,
\end{equation}
up to order-one coefficients. Symmetry protection therefore allows $\Delta\mathcal{E}$ to remain parametrically small for several e-folds, enabling violations of the classical bound $\mathcal{E}^{(\mathrm{cl})}$ of the magnitude shown in Fig.~1. In this way, the condition $\Gamma_{\rm dec}\ll H$ directly controls whether the inequality violations targeted by large-scale structure and 21\,cm surveys can survive to late times, see Appendix \ref{app:symmetry} for more details.

If fluctuations in the protected sector are subsequently transferred to the curvature perturbation---for example through modulated reheating, curvaton-like conversion, or symmetry-preserving interactions---they provide multiple correlated observables derived from the same underlying quantum state. These observables can then be used to construct the conditional variances appearing in Eq.~(\ref{eq:classical_bound}). In this case, the suppression of decoherence allows the conditional variance to fall below the classical noise floor, leading to a violation of the inequality.

Symmetry protection therefore plays a dual role in our framework. First, it provides a physically motivated mechanism for delaying the quantum-to-classical transition in a controlled subspace of the theory. Second, it naturally supplies the multiple observational kernels required to implement strengthened classicality inequalities with cosmological data. This makes symmetry-protected spectator sectors a natural arena in which inequality-based tests of primordial quantumness become observationally viable.

\section{Observational Pathways: Large-Scale Structure and 21\,cm Surveys}

The inequality-based tests proposed in this work require access to multiple correlated observables derived from the same primordial modes. While the cosmic microwave background provides a clean probe of primordial physics, it is fundamentally limited by cosmic variance and by the finite number of accessible modes. Large-scale structure (LSS) surveys and 21\,cm observations, by contrast, offer access to a vastly larger number of modes and multiple observational kernels, making them the natural arena for implementing classicality tests.

In LSS surveys, the primordial curvature perturbation $\zeta$ is mapped to late-time observables through scale- and redshift-dependent transfer functions and bias relations. Different tracers of the matter density field (e.g.\ galaxies, quasars, intensity mapping) and different redshift bins effectively provide distinct linear kernels acting on the same primordial modes. These kernels define the observables $X_i$ introduced in Eq.~(4), allowing the construction of cross-correlations and conditional variances required to test the classicality inequalities derived above. Multi-tracer techniques are particularly advantageous, as they reduce cosmic variance and enhance sensitivity to correlated primordial signals.

Tomographic 21\,cm surveys provide an even richer observational setting. Measurements of neutral hydrogen across a wide range of redshifts yield access to an enormous number of Fourier modes, potentially orders of magnitude beyond those available in the CMB. In particular, post-reionization intensity mapping and, ultimately, dark-ages 21\,cm observations probe the linear regime of structure formation over extended redshift intervals, enabling the construction of multiple, well-separated kernels acting on the same primordial perturbations.

A key advantage of the present framework is that the proposed inequalities involve only two-point correlators and conditional variances constructed from observable fields. As such, they can be implemented within existing LSS and 21\,cm data pipelines, subject to realistic modeling of late-time nonlinearities, bias, and foregrounds. While these effects modify the numerical values of the correlators and reduce the effective number of usable modes, they do not generically restore a positive classical probability representation once it has been violated. The inequalities therefore provide a robust diagnostic of the primordial origin of correlations, complementary to traditional searches for specific non-Gaussian templates.
\begin{figure*}[ht!]
    \centering
    \includegraphics[width=.8\linewidth]{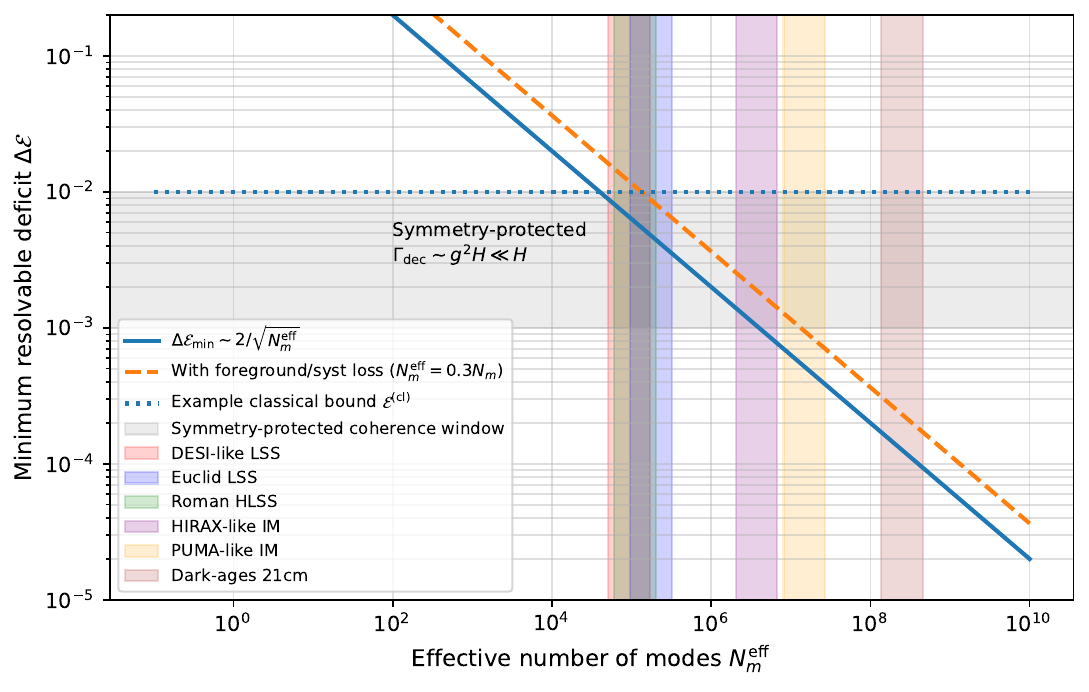}
\caption{Minimum resolvable deficit $\Delta\mathcal{E}$ of the conditional-variance statistic $\mathcal{E}=1-r^2$ as a function of the effective number of usable Fourier modes $N_m^{\rm eff}$. 
The solid curve shows the Gaussian scaling $\Delta\mathcal{E}_{\min}\simeq 2/\sqrt{N_m^{\rm eff}}$, while the dashed curve illustrates the impact of foregrounds and systematics, modeled as a reduction $N_m^{\rm eff}=0.3\,N_m$. 
The horizontal dotted line indicates an illustrative classical bound for local stochastic models with bounded noise $\mathcal{E}^{(\mathrm{cl})}=10^{-2}$, while the shaded horizontal band highlights the range $\Delta\mathcal{E}\sim10^{-3}$–$10^{-2}$ naturally populated by symmetry-protected spectator sectors, for which decoherence is parametrically suppressed ($\Gamma_{\rm dec}\sim g^2H\ll H$). 
Shaded vertical bands indicate representative ranges of $N_m^{\rm eff}$ for multi-tracer galaxy surveys (DESI, Euclid, Roman), post-reionization 21\,cm intensity mapping experiments (HIRAX- and PUMA-like), and an idealized dark-ages 21\,cm survey. 
Crossing the classical bound corresponds to falsification of a classical stochastic description of primordial fluctuations.}
 \label{fig:main}
\end{figure*}
To make the observational requirements explicit, consider two late-time observables (e.g.\ two tracers or redshift slices) modeled as $\delta_i(\mathbf{k}) = T_i(k)\zeta(\mathbf{k}) + n_i(\mathbf{k})$. The relevant classicality statistic can be written as
\begin{equation}
\mathcal{E}(k) \equiv 1 - r^2(k), \qquad
r \equiv \frac{P_{12}}{\sqrt{P_{11}P_{22}}},
\end{equation}
which is directly estimable from auto- and cross-spectra, see Appendix \ref{app:obs}. For Gaussian fields, error propagation yields $\sigma_{\mathcal{E}}\sim2/\sqrt{N_m^{\rm eff}}$ near $r\simeq1$, where $N_m^{\rm eff}$ denotes the effective number of usable Fourier modes after accounting for survey volume, scale cuts, and foreground losses. Resolving a deficit $\Delta\mathcal{E}$ therefore requires $N_m^{\rm eff}\gtrsim4/\Delta\mathcal{E}^2$.

Figure~\ref{fig:main} summarizes these scalings and illustrates the observational reach of different survey classes. Multi-tracer galaxy surveys with conservative linear-scale cuts achieve $N_m^{\rm eff}\sim10^5$, corresponding to sensitivity to percent-level deficits $\Delta\mathcal{E}\sim\mathcal{O}(10^{-2})$. Post-reionization 21\,cm intensity mapping experiments can reach $N_m^{\rm eff}\sim10^6$–$10^7$, enabling sensitivity to sub-percent deficits, while dark-ages 21\,cm tomography represents an ultimate benchmark with $N_m^{\rm eff}\gtrsim10^8$. These estimates demonstrate that inequality-based tests are naturally aligned with the mode-rich regimes targeted by LSS and 21\,cm surveys, and that forthcoming data can probe classicality bounds of direct relevance to the survival of primordial quantum coherence.

\section{Conclusions}

In this work we have addressed a foundational question in early-universe cosmology: whether primordial fluctuations provide empirical evidence for a quantum field theoretic description of inflation, beyond their successful phenomenological modeling. While inflationary perturbations are conventionally computed as quantum vacuum fluctuations, we emphasized that agreement with the scalar power spectrum alone does not experimentally establish their quantum origin, as classical stochastic models can reproduce these observables by construction.

We formulated this issue as a falsification problem by defining a precise classical stochastic hypothesis, based on locality, causality, and the existence of a positive probability distribution over initial conditions. We showed that this hypothesis implies a hierarchy of inequality constraints on observable correlators, and we identified strengthened classicality inequalities—formulated in terms of conditional variances—that need not be satisfied by quantum states generated during inflation. Violation of these inequalities therefore rules out a broad and well-defined class of classical alternatives, without relying on Bell inequalities or non-commuting measurement settings.

A key challenge for such tests is the rapid emergence of effective classicality during inflation due to squeezing and decoherence. We showed that this obstacle can be parametrically alleviated in the presence of symmetry-protected spectator sectors, in which conserved charges or approximate symmetries suppress decoherence and extend the window over which quantum coherence survives. These sectors naturally provide multiple correlated observables required to implement inequality-based tests.

We quantified the observational requirements for these tests by relating the resolvable deficit in the classicality inequalities to the effective number of usable Fourier modes. Using conservative assumptions, we showed that multi-tracer galaxy surveys are sensitive to percent-level violations, while next-generation 21\,cm intensity mapping experiments may probe the sub-percent regime. Dark-ages 21\,cm tomography represents an ultimate sensitivity benchmark, albeit with substantial astrophysical and instrumental challenges.

Our results reframe the question of primordial quantumness from an idealized Bell experiment to an observational consistency test aligned with realistic cosmological data pipelines. Rather than attempting to prove that the early universe was quantum in an axiomatic sense, the approach developed here provides a concrete and falsifiable route to excluding classical stochastic descriptions. As mode-rich surveys continue to expand the accessible volume of the universe, inequality-based tests offer a viable path toward empirically probing the quantum nature of the primordial universe.

\bibliographystyle{apsrev4-2}
\bibliography{references}

\appendix
\FloatBarrier
\begin{center}
\LARGE
{\bf Appendices}
\end{center}

\section{Observational Implementations and Scaling Estimates}
\label{app:obs}

In this Appendix we collect explicit definitions and scaling estimates for the inequality-based observables discussed in the main text, and outline their implementation in several observational settings. The purpose is not to provide a full survey forecast, but to establish order-of-magnitude requirements and to clarify how volume, tracer multiplicity, and large-scale access control sensitivity to classicality violations.

\subsection{Fiducial Inequality and Estimator}

We consider late-time observables $\delta_i(\mathbf{k})$ related linearly to the primordial curvature perturbation,
\begin{equation}
\delta_i(\mathbf{k}) = T_i(k)\,\zeta(\mathbf{k}) + n_i(\mathbf{k}),
\end{equation}
where $T_i(k)$ encodes transfer functions and bias, and $n_i$ denotes noise (shot noise, thermal noise, or residual foregrounds). The relevant power spectra are
\begin{equation}
P_{ij}(k) \equiv \langle \delta_i(\mathbf{k})\delta_j(-\mathbf{k})\rangle .
\end{equation}

A convenient classicality statistic is the conditional variance,
\begin{equation}
\mathrm{Var}(\delta_1|\delta_2)(k) = P_{11}(k) - \frac{P_{12}^2(k)}{P_{22}(k)}
= P_{11}(k)\bigl(1-r^2(k)\bigr),
\end{equation}
where
\begin{equation}
r(k) \equiv \frac{P_{12}(k)}{\sqrt{P_{11}(k)P_{22}(k)}}
\end{equation}
is the correlation coefficient. Defining
\begin{equation}
\mathcal{E}(k) \equiv 1-r^2(k),
\end{equation}
the classical hypothesis $H_{\mathrm{C}}$ implies $\mathcal{E}(k)\ge \mathcal{E}^{(\mathrm{cl})}(k)$, where $\mathcal{E}^{(\mathrm{cl})}$ is a classical noise floor fixed by the existence of a positive probability distribution.

\paragraph{On the classical noise floor $\mathcal{E}^{(\mathrm{cl})}$.}
The classical bound $\mathcal{E}^{(\mathrm{cl})}$ entering the inequality tests should be understood as the minimal residual conditional variance compatible with the classical hypothesis $H_{\mathrm{C}}$, given finite-resolution observables and local stochastic dynamics. While a formally arbitrary classical probability distribution could, in principle, reproduce arbitrarily strong correlations by engineering nonlocal fine-tuned initial conditions across modes, such constructions lie outside the scope of $H_{\mathrm{C}}$. Under the assumptions of locality, causality, and bounded noise implicit in $H_{\mathrm{C}}$, the residual conditional variance cannot be made arbitrarily small and defines a finite classical noise floor. The precise numerical value of $\mathcal{E}^{(\mathrm{cl})}$ is therefore survey- and observable-dependent and is not universal; however, the existence of any nonzero lower bound is sufficient for falsification. The sensitivity estimates presented in this work should accordingly be interpreted as determining which ranges of classical bounds can be tested by a given survey, rather than as predictions of a unique threshold.

\subsection{Estimator Variance and Mode Counting}

For Gaussian fields, the variance of the cross-spectrum estimator in a shell of width $\Delta k$ is
\begin{equation}
\mathrm{Var}\!\left[\hat P_{12}(k)\right]
\simeq \frac{1}{N_m(k)}\left(P_{12}^2(k)+P_{11}(k)P_{22}(k)\right),
\end{equation}
where the number of independent modes is
\begin{equation}
N_m(k)\simeq \frac{V}{2\pi^2}k^2\Delta k,
\qquad
k_{\min}\simeq \frac{2\pi}{V^{1/3}} .
\end{equation}

Propagating errors yields, for $r\simeq 1$,
\begin{equation}
\sigma_{\mathcal{E}}(k)\sim \frac{2}{\sqrt{N_m^{\mathrm{eff}}(k)}},
\end{equation}
where $N_m^{\mathrm{eff}}$ accounts for noise degradation and scale cuts, $N_m^{\mathrm{eff}}= f_{fg} N_m$. Resolving a deficit $\Delta\mathcal{E}$ therefore requires
\begin{equation}
N_m^{\mathrm{eff}}\gtrsim \frac{4}{(\Delta\mathcal{E})^2}.
\end{equation}

\subsection{Large-Scale Structure: Multi-Tracer Surveys}

In galaxy surveys, different tracers (e.g.\ LRGs, ELGs, quasars) provide distinct kernels $T_i(k)$ acting on the same primordial modes. Multi-tracer combinations reduce cosmic variance and increase the effective correlation coefficient $r$, thereby reducing $\mathcal{E}$. For DESI-like volumes $V\sim (5\text{--}10)\,h^{-3}\mathrm{Gpc}^3$ and conservative linear cuts $k_{\max}\sim 0.05$--$0.1\,h\,\mathrm{Mpc}^{-1}$, one typically obtains
\begin{equation}
N_m^{\mathrm{eff}}\sim 10^3\text{--}10^4 ,
\end{equation}
allowing sensitivity to $\Delta\mathcal{E}\sim 10^{-2}$ in idealized multi-tracer regimes. This is consistent with order-unity sensitivity to primordial long--short couplings achieved in multi-tracer PNG forecasts.

\subsection{21\,cm Intensity Mapping}

21\,cm intensity mapping surveys provide three-dimensional tomographic access to the matter field over large volumes and redshift ranges. Distinct redshift slices act as independent kernels $T_i(k,z)$, enabling conditional-variance tests using cross-correlations between slices. For next-generation experiments with $V\sim 10^2\,h^{-3}\mathrm{Gpc}^3$ and $k_{\min}\sim 10^{-3}\,h\,\mathrm{Mpc}^{-1}$, one can reach
\begin{equation}
N_m^{\mathrm{eff}}\sim 10^4\text{--}10^5 ,
\end{equation}
in foreground-clean regions, corresponding to potential sensitivity to $\Delta\mathcal{E}\sim 10^{-2}$--$10^{-3}$, subject to foreground and calibration control.

\subsection{Dark-Ages 21\,cm Tomography}

Dark-ages 21\,cm observations offer access to an enormous number of linear modes across a wide redshift range ($30\lesssim z\lesssim 100$). In an ideal cosmic-variance-limited scenario, one can achieve
\begin{equation}
N_m^{\mathrm{eff}}\gtrsim 10^6 ,
\end{equation}
allowing sensitivity to $\Delta\mathcal{E}\sim 10^{-3}$ or smaller. However, nonlinear secondary contributions and foreground contamination are expected to dominate the error budget unless modeled at the percent level, and therefore this scenario should be regarded as an ultimate sensitivity benchmark rather than a near-term forecast.

\subsection{Role of Tracer Multiplicity}

With $N_t$ tracers, the conditional variance of one tracer given all others is
\begin{equation}
\mathrm{Var}(\delta_1|\{\delta_{a\neq 1}\})
= P_{11}-\mathbf{P}_{1\perp}\,\mathbf{P}_{\perp\perp}^{-1}\,\mathbf{P}_{\perp 1},
\end{equation}
which decreases as tracer diversity and signal-to-noise improve. This generalized conditional variance is the natural object entering inequality tests in multi-tracer analyses, and provides a direct route to beating cosmic variance in classicality tests, analogously to multi-tracer PNG measurements.

\subsection{Interpretation}

The estimates above demonstrate that inequality-based tests of primordial classicality are naturally aligned with the observational regimes already targeted by LSS and 21\,cm surveys. While near-term galaxy surveys are sensitive to percent-level deficits in $\mathcal{E}$, next-generation 21\,cm experiments may access substantially smaller values. These scalings justify the focus on mode-rich, multi-kernel observables as the appropriate arena for falsifying classical stochastic descriptions of primordial fluctuations.

\section{Symmetry-Protected Sectors and the Size of Observable Inequality Violations}
\label{app:symmetry}

In this Appendix we provide additional detail on the role of symmetry-protected spectator sectors in controlling the magnitude of observable violations of classicality inequalities. The purpose is not to construct a specific model, but to justify parametrically the range of deficits $\Delta\mathcal{E}$ discussed in the main text and illustrated in Fig.~1.

\subsection{Decoherence Rate and Symmetry Suppression}

For a generic light spectator field during inflation, interactions with other fields and with metric perturbations induce decoherence at a rate of order the Hubble scale. However, when the spectator sector is protected by an exact or approximate continuous symmetry, the leading interactions are constrained by selection rules or derivative couplings. As a result, the effective coupling $g$ controlling interactions with the environment is parametrically suppressed.

In such cases, the decoherence rate can be estimated on dimensional grounds as
\begin{equation}
\Gamma_{\rm dec} \sim g^2 H ,
\end{equation}
where $H$ is the Hubble parameter during inflation. Values $g\ll1$ arise naturally for Goldstone-like modes, axion-like fields, or phases associated with conserved charges, without requiring fine-tuning of parameters.

\subsection{Connection to the Conditional-Variance Deficit}

Decoherence acts to suppress correlations between different observational kernels and therefore drives the conditional-variance statistic
\begin{equation}
\mathcal{E} \equiv 1 - r^2
\end{equation}
toward its classical value. Over a time interval $\Delta t$ after horizon exit, one expects schematically
\begin{equation}
\Delta\mathcal{E}_{\rm dyn} \sim \Gamma_{\rm dec}\,\Delta t ,
\end{equation}
up to coefficients of order unity that depend on the precise interaction structure.

During inflation, the relevant interval before decoherence or conversion to curvature perturbations is finite and can be parametrized as $\Delta t \sim N/H$, where $N$ is the number of e-folds over which coherence is preserved. Combining these estimates yields
\begin{equation}
\Delta\mathcal{E}_{\rm dyn} \sim g^2 N .
\end{equation}

This relation provides a direct link between symmetry protection, decoherence, and the magnitude of observable inequality violations.

\subsection{Natural Range of $\Delta\mathcal{E}$}

For symmetry-protected sectors, values of the effective coupling in the range $g\sim10^{-1}$–$10^{-2}$ are natural, corresponding to weak derivative interactions or loop-suppressed couplings. For coherence lasting over $N\sim\mathcal{O}(1$–$10)$ e-folds, this leads to
\begin{equation}
\Delta\mathcal{E}_{\rm dyn} \sim 10^{-2}\text{–}10^{-3} .
\end{equation}

This range does not require extreme sequestering of the spectator sector and represents the generic regime in which symmetry protection has an observable impact. Much smaller values $\Delta\mathcal{E}\ll10^{-3}$ would require both exceptionally small couplings and unusually long coherence times, and are therefore increasingly model-dependent. Conversely, larger values $\Delta\mathcal{E}\gtrsim10^{-1}$ are generically produced even in the absence of symmetry protection and do not provide discriminatory power between classical and quantum origins.

\subsection{Relation to Observational Sensitivity}

The range $\Delta\mathcal{E}\sim10^{-2}$–$10^{-3}$ therefore represents the natural overlap between dynamical viability and observational sensitivity. As shown in Fig.~1, percent-level deficits are accessible to multi-tracer galaxy surveys, while sub-percent deficits are within reach of next-generation 21\,cm experiments. Symmetry protection thus does not introduce an additional observational requirement, but determines whether physically motivated models can populate the region of parameter space that surveys are capable of probing.

In this sense, symmetry-protected sectors provide a controlled mechanism by which violations of classicality inequalities of observable magnitude can survive inflationary evolution and be imprinted on late-time cosmological observables.

\subsection{A Toy Model with a Shared Conserved Charge}

To make the symmetry-protection mechanism explicit, it is useful to consider a simple toy model in which quantum coherence is protected by a shared conserved charge, while inflation itself is driven by an independent inflaton sector. The purpose of this model is illustrative: it demonstrates how symmetry protection arises at the level of field variables, without relying on detailed model building.

We consider an inflaton field $\phi$, responsible for slow-roll inflation, and two complex spectator fields $\Phi_1$ and $\Phi_2$ carrying a common global $U(1)$ charge. Working on an FRW background with metric $ds^2=-dt^2+a^2(t)d\mathbf{x}^2$, a minimal action is
\begin{eqnarray}
S &=&\int d^4x\sqrt{-g}\Big[
\frac{M_{\rm Pl}^2}{2}R
-\frac12(\partial\phi)^2
-V(\phi) \nonumber\\
&-&\sum_{a=1}^2 |\partial\Phi_a|^2
-U(|\Phi_1|,|\Phi_2|)
-\lambda|\Phi_1|^2|\Phi_2|^2
\Big],
\end{eqnarray}
where the inflaton potential $V(\phi)$ satisfies standard slow-roll conditions and the $\Phi$ sector is assumed to be a spectator, with subdominant energy density. The potential $U$ depends only on the moduli $|\Phi_a|$, ensuring that the global $U(1)$ symmetry
\begin{equation}
\Phi_a \rightarrow e^{i\alpha}\Phi_a, \qquad a=1,2,
\end{equation}
is exact.

A convenient parametrization is obtained by writing the complex fields in polar form,
\begin{equation}
\Phi_a(x)=\frac{1}{\sqrt{2}}\rho_a(x)e^{i\theta_a(x)}.
\end{equation}
The symmetry acts by a common shift of the phases $\theta_a\rightarrow\theta_a+\alpha$, implying the existence of a conserved Noether charge in the $\Phi$ sector.

It is then natural to decompose the phase degrees of freedom into a \emph{charged} (symmetry-protected) combination and an orthogonal relative phase,
\begin{equation}
\Theta \equiv \frac{\rho_1^2\theta_1+\rho_2^2\theta_2}{\rho_1^2+\rho_2^2},
\qquad
\Delta \equiv \theta_1-\theta_2.
\end{equation}
The variable $\Theta$ transforms under the global $U(1)$ and is directly tied to the conserved charge, while $\Delta$ is invariant. As a result, $\Theta$ couples to other degrees of freedom only through derivative interactions or symmetry-breaking effects, whereas $\Delta$ and the radial modes $\rho_a$ are generically unprotected.

This structure has direct consequences for decoherence. Environmental interactions and metric perturbations can readily entangle the unprotected degrees of freedom, but decoherence of the charged phase $\Theta$ is suppressed by the symmetry. In particular, the effective coupling controlling decoherence of $\Theta$ is parametrically small, leading to a rate $\Gamma_{\rm dec}\sim g^2H$ with $g\ll1$. Quantum coherence in the $\Theta$ sector therefore survives for several e-folds after horizon exit.

From the perspective of the classicality inequalities introduced above, the relevance of this toy model is that observables constructed from the charged phase inherit strong correlations protected by the conserved charge. When different observational kernels probe the same $\Theta$ fluctuations, their cross-correlations remain large, keeping the conditional-variance statistic $\mathcal{E}=1-r^2$ parametrically small. This provides a concrete realization of how symmetry protection allows violations of the classical stochastic bounds to survive inflationary evolution with observable magnitude.

The physical reason why coherence survives in the charged phase $\Theta$ can be understood as a direct consequence of charge conservation. Decoherence arises when a degree of freedom becomes entangled with an environment in a way that allows the environment to ``measure'' its state. For generic spectator fields during inflation, interactions with other fields or with short-wavelength metric perturbations continuously record information about field amplitudes, leading to rapid suppression of phase coherence on super-Hubble scales.

In contrast, the charged phase $\Theta$ is tied to a conserved Noether charge. Any interaction capable of distinguishing different values of $\Theta$ must therefore violate the $U(1)$ symmetry or involve derivatives of $\Theta$, which are suppressed on super-Hubble scales. As a result, the environment has no efficient channel through which to acquire information about the absolute value of $\Theta$. In this sense, $\Theta$ behaves as a ``protected'' collective coordinate: while its conjugate momentum is diluted by cosmic expansion, the phase itself is not monitored by the environment, and quantum superpositions of $\Theta$ are not rapidly decohered.

Equivalently, in the language of open quantum systems, the reduced density matrix for $\Theta$ remains approximately diagonal in the charge basis rather than in the field-amplitude basis. Since the charge is conserved, the interaction Hamiltonian cannot generate rapid phase-dependent entanglement with environmental degrees of freedom. Decoherence of $\Theta$ therefore proceeds only through symmetry-breaking effects or higher-derivative interactions, both of which are parametrically suppressed. This leads to a decoherence rate $\Gamma_{\rm dec}\sim g^2H$ with $g\ll1$, in contrast to the generic case where $\Gamma_{\rm dec}\sim H$.

The consequence for observable correlations is direct. Observables constructed from fluctuations of the charged phase inherit this protection: different observational kernels probing the same $\Theta$ fluctuations remain strongly correlated over several e-folds after horizon exit. While unprotected modes are driven toward a classical stochastic description, the protected sector retains phase coherence long enough to imprint correlations that violate the classicality bounds discussed in Sec.~III. In this way, symmetry protection provides a concrete dynamical mechanism allowing violations of the classical stochastic hypothesis to survive inflationary evolution with observable magnitude.

\end{document}